# Design-Space Exploration and Optimization of an Energy-Efficient and Reliable 3D Small-world Network-on-Chip

Sourav Das, *Student Member, IEEE*, Janardhan Rao Doppa, *Member, IEEE,* Partha Pratim Pande, *Senior Member, IEEE,* Krishnendu Chakrabarty, *Fellow, ACM and IEEE*

*Abstract*— A three-dimensional (3D) Network-on-Chip (NoC) enables the design of high performance and low power many-core chips. Existing 3D NoCs are inadequate for meeting the ever-increasing performance requirements of many-core processors since they are simple extensions of regular 2D architectures and they do not fully exploit the advantages provided by 3D integration. Moreover, the anticipated performance gain of a 3D NoC-enabled many-core chip may be compromised due to the potential failures of through-silicon-vias (TSVs) that are predominantly used as vertical interconnects in a 3D IC. To address these problems, we propose a machine-learning-inspired predictive design methodology for energy-efficient and reliable many-core architectures enabled by 3D integration. We demonstrate that a small-world network-based 3D NoC (3D SWNoC) performs significantly better than its 3D MESH-based counterparts. On average, the 3D SWNoC shows 35% energy-delay-product (EDP) improvement over 3D MESH for the PARSEC and SPLASH2 benchmarks considered in this work. To improve the reliability of 3D NoC, we propose a computationally efficient spare-vertical link (sVL) allocation algorithm based on a state-space search formulation. Our results show that the proposed sVL allocation algorithm can significantly improve the reliability as well as the lifetime of 3D SWNoC.

## I. INTRODUCTION

THREE-dimensional (3D) ICs are capable of achieving better performance, functionality, and packaging density compared to their traditional planar counterparts [1][2]. On the other hand, network-on-chip (NoC) enables integration of large numbers of embedded cores in a single die. 3D NoC architectures combine the benefits of these two new paradigms to offer an unprecedented performance gain [2][3]. With freedom in the third (vertical) dimension, NoC architectures that were previously impossible or prohibitive due to wiring constraints in planar ICs are now realizable in 3D NoC, and many 3D implementations can outperform their 2D counterparts. However, existing 3D NoC architectures predominantly follow straightforward extensions of regular 2D NoC designs, which do not fully exploit the advantages provided by the 3D integration technology [3]. Another challenge is that, the anticipated performance gain of 3D NoC-enabled many-core chips may be compromised due to potential failures of the through-silicon-vias (TSVs) used as vertical interconnects. TSVs in a 3D IC fail due to voids, cracks, and different kinds of fabrication challenges [4]. Additionally, the workload induced stress increases the resistance of the TSVs, which leads to different mean-time-to-failure (MTTF) for different TSVs [5][6].

The main focus of this paper is to explore and consequently establish performance-energy-reliability trade-offs for 3D small-world NoC (SWNoC) [7][8]. To this end, we make the following contributions:

- We consider the design space of 3D SWNoC architectures, where the vertical connections predominantly work as long-range shortcuts for small-world networks. 3D SWNoC architectures (as shown in Fig. 1) help with both energy-efficiency (small average path length) and reliability (average path length grows insignificantly due to link failures). This is the first work where we exploit the advantages of 3D integration to design a power-law based small-world network-enable 3D NoC architecture.

- The design space of a 3D SWNoC is combinatorial in nature. Hence, we leverage machine-learning techniques to intelligently explore the design space to optimize the placement of both planar and vertical communication links for high performance and energy efficiency.

- We consider spare vertical link (sVL) allocation to improve the reliability of the 3D NoC. This is another combinatorial optimization problem, where we do not know the cost function. We can experimentally compute the quality (or cost) of a solution by running a simulation. We solve this problem using a state-space search formulation, where the simulations guide the search process. We leverage the structure of the problem and domain knowledge of the 3D SWNoC to efficiently produce a sVL allocation that can significantly improve the reliability of the 3D NoC.

- We perform a comprehensive experimental study by using several PARSEC and SPLASH2 benchmarks to evaluate the proposed optimized 3D NoC architecture, and spare-vertical link (sVL) allocation schemes. We show that the proposed 3D SWNoC outperforms the state-of-the-art NoC architectures for all benchmarks considered in this work. We

This work was supported in part by the US National Science Foundation (NSF) grants CNS 1564014, CCF-0845504, CNS-1059289, and CCF-1162202, and Army Research Office grant W911NF-12-1-0373.

Sourav Das, Janardhan Rao Doppa and Partha Pratim Pande are with the School of Electrical Engineering and Computer Engineering, Washington State University, Pullman, WA, USA. e-mail: {sdas, jana, pande}@eecs.wsu.edu.

Krishnendu Chakrabarty is with the Department of Electrical and Computer Engineering, Duke University, Durham, NC, USA. email:krish@duke.edu

show the effectiveness of our greedy sVL allocation method by comparing its computation time and solution quality with those obtained via exhaustive search. Finally, we also demonstrate the soundness of domain knowledge used for pruning the search space for sVL allocation for 3D SWNoC.

The rest of the paper is organized as follows: Section II describes related work. We describe the 3D SWNoC architecture in Section III and our 3D NoC design optimization methodology based on machine-learning in Section IV. In Section V, we explain our sVL allocation algorithms. We present the experimental setup, performance of 3D SWNoC and related analysis in Section VI. The performance evaluation of sVL allocation algorithms is described in section VII. Finally, section VIII concludes the paper by summarizing the salient features of this work.

## II. RELATED PRIOR WORK

We categorize the prior work on 3D NoC design as follows:

### A. 3D NoC Architectural Space

Most of the existing 3D NoC architectures are based on a conventional mesh topology [9][10][11]. However, it is well-known that mesh-based architectures suffer from high network latency and energy consumption due to their multi-hop communication links. To exploit the reduced distance along the vertical dimension of 3D IC, an NoC-bus hybrid architecture was proposed in [12]; it uses Dynamic Time Division Multiple Access (dTDMA) to reduce the network latency. To reduce energy consumption of the system, the 3D Dimensionally Decomposed (DimDe) NoC router architecture [13] was developed. In an NoC, the largest percentage of energy is consumed by the routers and energy consumption increases non-linearly with the number of input ports. Hence to reduce the energy consumption and the number of input ports, an improved 3D NoC router architecture was developed [14]. All of these architectures have buses in the Z-dimension; and hence, with increasing network size, they are subject to traffic congestion and high latency under high traffic injection loads.

The Sunfloor 3D was developed for synthesizing application specific 3D NoCs [15]. The design and synthesis of application-specific 3D NoC architectures was also investigated [16][17]. Later, a more general-purpose 3D NoC was proposed in [18] using an ILP-based algorithm to insert long-range links to develop low diameter and low radix architecture. However, the reduction in energy consumption was found to be limited.

Photonic interconnects offer high bandwidth and low power for future many-core chip design. A number of hybrid 3D/photonic NoC architectures have been designed [2][19]. However, on-chip photonics still suffers from performance variation due to thermal issues [20]. In addition, the challenges of integrating two emerging paradigms, namely 3D IC and silicon nano-photonics, are yet to be adequately addressed.

In this work, we focus on designing a robust 3D NoC architecture that combines the benefits of 3D IC and the robustness of the SW network. We present a detailed design methodology and an optimization algorithm based on machine-learning for developing energy-efficient 3D NoC architectures. We also perform comparative performance analysis with respect to conventional 3D MESH and other irregular architectures. We show that our proposed architecture performs better than other existing 3D NoCs when provided with the same amount of physical resources.

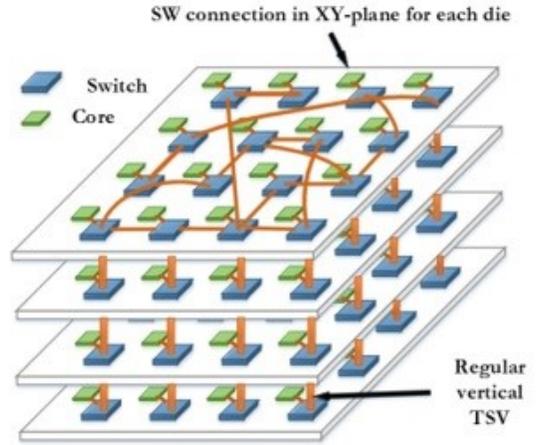

Fig. 1: Conceptual view of 3D SWNoC with TSV-based vertical link. For simplicity, only one logical XY-plane SW connection is shown.

### B. Reliability Analysis

The performance of TSV-based 3D ICs degrades due to TSV failure. To overcome such performance penalties arising from TSV failure, researchers have investigated spare TSV (sTSV) allocation and sharing scheme for 3D ICs [4][21]. The initial sharing algorithm was developed based on the idea of utilizing one extra TSV for each of the functional TSVs [21]. However, the reliability improvement comes at the expense of double TSV count and significant area overhead.

To avoid the 100% TSV area overhead, researchers have proposed several TSV sharing schemes (e.g., 3:2, 4:1, and so on) [22][23]. The main idea is to share spare TSV/s among a group of functional TSVs to compensate for performance penalty due to possible TSV failure within that particular group. However, depending on the sharing scheme, a significant amount of encoder and decoder logic circuits are necessary to shift the signals and select the correct spare TSV, which introduces additional delay and power consumption. In addition, the delay for each TSV can be different depending on the location of the failed TSV, which may result in timing violation. To address the varying delay for each TSV, a group-based 6-TSV placement scheme for 4 functional TSVs was proposed to improve the reliability of a 3D DRAM [24]. With the help of a switchbox-based design for each group, correct signals were selected and transferred for functional TSVs. The main advantage was that the same amount of delay was incurred by every TSV in the box. However, this advantage comes at the expense of 50% area overhead and significant power consumption from the switch-boxes. Similarly, researchers have developed a block-based redundancy architecture and used signal-shifting techniques for fault tolerance [25][26]. In this case, if any TSV fails, then the signal shifts towards the redundant ones. The signal shifting technique can tolerate one TSV failure. To improve fault tolerance for more than one TSV failure, a crossbar-based redundant TSV architecture was developed in [27]. However, the number of redundant TSVs increases significantly in this case. TSV resource sharing algorithms, which can be selectively applied depending on the sharing granularity and design complexity were also developed. Word-level and bit-level TSV sharing was formulated as a constrained clique-partitioning problem and efficient algorithms were designed to solve it. However, these algorithms do not scale for large-scale design problems.

All the above-mentioned TSV sharing schemes improve the performance of 3D ICs and hence, the overall reliability as well. However, the allocation of spare TSVs for 3D NoCs need to consider additional constraints arising from the physical NoC design perspective. In a 3D NoC, TSVs are placed in a bundle to enable a single vertical link (VL). Depending on the physical placements of routers and cores of 3D NoCs, these VLs maintain considerable physical distance between them. Hence, sharing TSVs among these VLs is not feasible due to the physical design and timing constraints. In addition, if one TSV fails in a VL, then the achievable performance of the whole link is affected, which in turn degrades the overall NoC performance. Hence, we focus on spare-vertical link (sVL) allocation instead of individual spare TSVs.

In this work, to analyze the reliability issues of 3D NoC, we evaluate the performance of a 3D NoC with workload-induced VL failure. We perform a comparative reliability analysis study of different 3D NoCs with VL failure. Then we describe the sVL allocation mechanism for a 3D NoC. We formulate sVL allocation as an optimization problem to minimize the performance penalty due to TSV-based VL failure. At the same time, we focus on improving the lifetime of the overall 3D NoC. We demonstrate two different algorithms viz. *greedy* and *exhaustive* search to allocate the sVLs in a 3D SWNoC to compensate for the performance penalty due to VL failure. We also compare the performance of both algorithms in terms of quality of the solution and computation time. We show that based on the domain knowledge of a 3D NoC, we can develop computationally efficient algorithms whose performances are similar to *exhaustive* search, a naïve approach.

## III. 3D NoC ARCHITECTURE DESIGN

In this section, we first describe the design of a small-world network based 3D NoC architecture. Next, we discuss the main challenges for developing an energy-efficient 3D NoC and the motivation for a machine-learning based optimization algorithm.

### A. Problem Description

The goal of an on-chip communication system design is to transmit data with low latencies and high throughput using the least possible power and resources. In this context, design of SW network-based NoC architectures [7] is a notable example. It has been shown that either by inserting long-range shortcuts in a regular mesh architecture to induce a SW effect or by adopting a power-law based SW connectivity, it is possible to achieve significant performance gain and lower energy consumption compared to traditional multi-hop mesh networks [7][8]. In this work, we advocate that the concept of small-worldness should be adopted in 3D NoCs too. Specifically, the VLs in 3D NoC should enable the design of long-range shortcuts necessary for a SW network. However, the appropriate placement of the planar and the long-range links along the vertical dimension is crucial for maximizing the performance benefits. Hence, our goal is to optimize the placement of the planar and the vertical links in a 3D NoC where the overall interconnection architecture follows the small-world connectivity, and improves the network latency and power consumption per message.

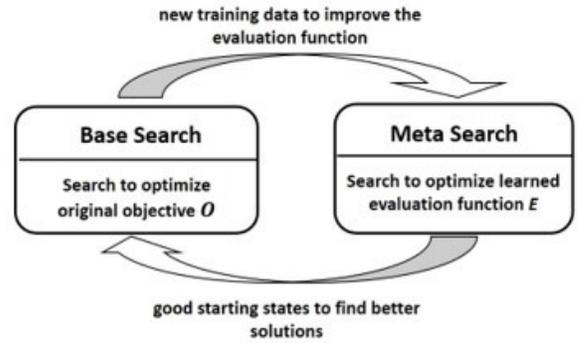

Fig. 2: High-level overview of the optimization algorithm.

### B. Small-world (SW) Network

An SW network lies in-between a regular, locally interconnected mesh network and a completely random Erdös-Rényi topology. SW graphs have a very short average path length, defined as the number of hops between any pair of nodes. The average shortest path length of SW graphs is bounded by a polynomial in log ($N$), where $N$ refers to the number of nodes; this property makes SW graphs particularly interesting for efficient communication with minimal resources [28][29]. To develop small-world network, we follow the power-law based connectivity [28][29]. The probability ($p$) of having a direct link between nodes in a SW network varies exponentially with the link length ($\ell$), i.e., $p(\ell) \propto \ell^{-\alpha}$ where, the parameter $\alpha$ governs the nature of connectivity, e.g., a larger $\alpha$ means a locally connected network with a few, or even no long-range links. By the same token, a zero value of $\alpha$ generates an ideal SW network following the Watts-Strogatz model [28][29][30]– one with long-range shortcuts that are virtually independent of the distance between the cores. In this work, we also determine the best suited connectivity parameter, $\alpha$, for a small-world network enabled 3D SWNoC architecture (experimentally determined and analyzed in section VI. B), which has been shown to produce the most energy efficient and high performance 3D SWNoC.

### C. Development of 3D SWNoC

Starting from a power-law based connectivity, we attempt to optimize the location of the planar links and the VLs to achieve lower latency and energy consumption. We define an objective function $O$ called communication cost, which combines the NoC performance metrics, namely the network latency and energy consumption per message. Optimizing the communication cost ensures lower average hop count and improvement in the network performance in terms of both latency and energy consumption. However, the space of physically feasible SW based 3D NoC designs $D$ is combinatorial in nature and our goal is to find the design $d \epsilon D$ that minimizes $O$. One could employ search algorithms such as hill-climbing and simulated annealing, which are very popular in the design community for this task. However, we leverage machine-learning techniques that have been shown to improve the performance of these search algorithms by intelligently exploring the design space [31][32]. This optimization process is undertaken before the actual NoC implementation.

### IV. NoC Optimization based on Machine-learning

We employ an online learning algorithm called STAGE [31], which was originally developed to improve the performance of local search algorithms (e.g., hill climbing) with random-restarts for combinatorial optimization problems. The high-level conceptual idea of the algorithm is shown in Fig. 2. The key insight behind STAGE is to leverage some extra features $\phi(d) \in R^m$ ($m$ is the number of features) of the optimization problem to learn an improved evaluation function $E$ that can estimate the promise of a design $d$ as a starting point for the local search procedure $A$. It employs $E$ to intelligently select promising starting states that will guide $A$ towards significantly better solutions. Past work in the search community concluded that many practical optimization problems exhibit a "globally convex" or "big valley" structure, where the set of local optima appear to be convex with one global optimum in the center [32]. The main advantage of STAGE over popular algorithms such as simulated annealing and Integer Linear Programming (ILP) is that it tries to learn the solution space structure, and uses this information in a clever way to improve both convergence time and the quality of the solution. This aspect of STAGE is very advantageous for large system sizes to improve the design-validate cycle before mass manufacturing and for dynamically adapting the designs for new application workload. To the best of our knowledge, this is the first work that applies STAGE to an NoC design optimization problem. Algorithm 1 provides the pseudocode for our NoC optimization technique.

**Challenges:** The main challenges in applying STAGE to our 3D NoC design optimization problem are as follows: 1) We need to define additional features of the optimization problem that can be exploited to learn improved evaluation functions for efficient design space exploration. We provide these features for 3D NoC designs (Table 1), but they can be adapted to other types of NoC designs as well; 2) Defining appropriate search spaces by leveraging the domain knowledge can potentially improve the effectiveness of the STAGE algorithm. We need to identify good starting state distribution (subset of initial 3D NoC design solutions) and search operators (actions to get successor states from a given state) to navigate the design space. We have explored γ-greedy for starting state distribution with the hope of improving over random starting state distribution (see "Starting States and Successor Function" below); and 3) We need to find a good knowledge representation for the evaluation function E that is expressive, can be trained quickly, and allows to make fast predictions. We picked regression trees as it satisfies all the requirements (see "Regression Learner" below).

#### A. Instantiation for 3D NoC Optimization

In this section, we provide all the details needed to apply the STAGE algorithm to our 3D NoC optimization problem.

**Design Space:** Our design space depends on a set of network resources, which are given as input to the optimization algorithm. These resources are defined as follows. 1) Cores (*C*): A set of all cores $C = \{C_1, C_2,...,C_N\}$, where $N$ is total number of cores. We assume that every core is connected to at least one router; 2) Planar Dies (*P*): A set of all dies $P$. For $N = 64$, we consider four dies with each die containing 16 cores. For core placement, we follow a greedy algorithm to minimize $(f_{ij}*d_{ij})$,

---

**Algorithm 1: NoC Design Optimization via STAGE Algorithm**

1: **Input:** $D$ = Design space,
   $O$ = cost function,
   $(I,S)$ = initial state and successor generation functions,
   $C$ = network constraints,
   $\phi$ = feature function for NoC design,
   $A$ = local search procedure,
   $R$ = regression learner,
   $MAX$ = maximum iterations,

2: **Output:** $d_{best}$, the best NoC design

3: **Initialization:** initialize evaluation function $E$, training set $Z$, initial design $d_0$, $O_{best} = O(d_0)$, and $d_{best} = d_0$

4: **Repeat:**

5:    **Base search**: From $d_0$, run the search procedure $A$ guided by $O$ until a local optima is reached, leading to a search trajectory ($d_0, d_1,..., d_T$).

6:    **Generate training data**: For each design $d_i$ on the search trajectory, add ($\phi(d_i)$, $y_i$) to Z, where $y_i$ is the best value along the search trajectory.

7:    **Re-train** $E$: $E = R(Z)$.

8:    **Meta search**: From $d_T$, run the search procedure $A$ guided by $E$ until a local optima is reached to produce the best predicted starting state $\hat{d}$.

9:    **Next starting state**: If $\hat{d} = d_T$ (no search progress), set $d_0$ using $I$. Otherwise, set $d_0 = \hat{d}$.

10:    **Update** $O_{best}$ and $d_{best}$ if $y^* < O_{best}$, where $y^*$ is the best value encountered during base search and meta search.

11: **Until** $MAX$ iterations or convergence.

12: **Return** best design $d_{best}$.

---

where $f_{ij}$ and $d_{ij}$ are the communication frequency and Cartesian distance between the cores respectively. In this step, we form clusters with 16 cores in each die; 3) Link Distribution (*L*): The link length distribution $L = \{l_1, l_2,...,l_k\}$, where $k$ depends on the size and topology of the network; $l_i$'s are determined based on the SW connectivity parameter $\alpha$. For higher values of $\alpha$, $l_k$ decreases; and 4) Communication Frequency (*F*): The communication frequency among different cores $F = \{f_{ij} \mid 1 \leq i, j \leq N, i \neq j\}$. We assume that $F$ for each application is given as an input to perform application-specific network optimization.

The set of all physically realizable SW NoC designs with the given link distribution $L$ forms our design space.

**Objective Function *O*:** We define $O$ as the communication cost of the given 3D NoC, which is the product of hop count, frequency of communication, and link length summed over every source and destination pair, i.e.,

$$O = \sum_{i=1}^{N} \sum_{j=1, i \neq j}^{N} (r * h_{ij} + d_{ij}) * f_{ij} \qquad (1)$$

where $f_{ij}$, and $d_{ij}$ are defined as above; $h_{ij}$ is hop count between $i$- and $j$-th node, and $r$ denotes the number of router stages. From a practical point-of-view, $r$ is the number of cycles a message spends inside a router to move from input to output port. An NoC design with low $O$ will have low latency and energy consumption, and hence, low energy-delay-product (EDP).

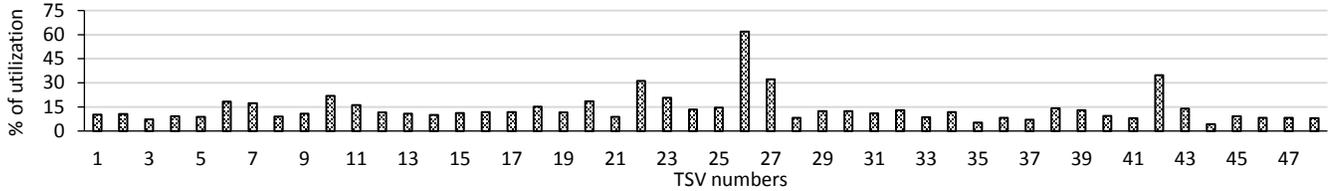

Fig. 3: Non-homogeneous VL utilization pattern of the 3D SWNoC for the CANNEAL benchmark. The region between 2nd and 3rd dies denoted by VLs numbering 17~32, and carry 45% of the total vertical-link traffics of the 4 die 3D system.

**Network Constraints:** To explore only physically feasible 3D NoC designs, we enforce some constraints on the placement of VLs and router configurations. If TSVs are considered as the VLs, we only allow placing them point-to-point (regularly) between the routers. Such constraints may put additional limits on the performance of NoC designs. However, efficient optimization can overcome such limitations. The SW network has an irregular connectivity. Hence, the number of links connected to each router is not constant. For fair comparison between our SW network and 3D MESH, we assume that both of them use the same average number of connections, $<k_{avg}>$ per router. This also ensures that the 3D SW NoC does not introduce additional links compared to a 3D MESH. For a 64-core system, $<k_{avg}>$ is 4.5 considering all the routers, including the peripheral ones. In addition, the maximum connectivity per node, $<k_{max}>$, is set to be 7 for the SW network as found in [33].

**Starting States and Successor Function:** For starting states, we randomly generate a SW network that satisfies the network constraints. The successor function $S$ takes a network as input and returns a set of next states, and allows the search procedure to navigate the NoC design space. $S$ generates one candidate state for each link connecting two nodes in the input network. It simply removes that link and places a link with the same length between two nodes in the network that are not directly connected.

The STAGE algorithm can benefit if we can specify the starting state distribution using some domain knowledge. Therefore, we also consider a starting-state distribution, named $\gamma$ -*Greedy*. We formulate the starting state (design) construction as a sequential decision-making task, where we select the next link to be placed at each step. In $\gamma$-*Greedy* distribution, we select a link greedily with probability $\gamma$ based on communication frequency and a random link with probability $(1-\gamma)$. We start with $\gamma=1$ (completely greedy) and gradually reduce $\gamma$ to increase the randomness.

**Local Search Procedure** $A$**:** We employed a stochastic hill-climbing procedure, where the next states are sampled stochastically.

**Feature Function** $\phi$**:** The main challenge in adapting STAGE to our NoC domain is to define a set of features $\phi$ for each network that can drive the learner. We divide the whole network into several overlapping subgraphs or regions, and define a set of features that can be categorized into three types: 1) *Average hop count* ($h$), which calculates the average hop count for each region or sub-network; 2) *Weighted communication* which is defined as the sum of the products of hop count and communication frequency over all source-destination pairs for a particular hop count ($\sum_{i=1}^{N}\sum_{j=1,j\neq i}^{N} f_{ij} * h_k$). The highest value of $k$ depends on the network size and topology. If the value of this feature is small, it indicates that highly communicating cores are placed in the same neighborhood; and 3) *Clustering coefficient* ($C_c$), which captures the connectivity of one core with its neighbors [34]. While the hop count takes into account mainly long-range communication, the clustering coefficient focuses more on local connectivity among the immediate neighbors. We found these features to sufficiently capture the network characteristics, efficient to compute, and allow learning highly accurate evaluation function, $E$.

TABLE I: FEATURE DESCRIPTION

| Feature Type | Feature Count |
|---|---|
| Avg. hop count for nine overlapping regions | 9 |
| Weighted communication ($f_{ij}*h_k$) considering maximum hop count for N=64 | 8 |
| Avg. Clustering coefficient ($C_c$) for four planar dies | 4 |

In this work, for $N$ = 64 cores, we divide the whole network into nine regions. For each region, we consider average hop counts as the features. In addition, the initial network has the highest hop count of eight, and hence we require eight features for weighted communication cost. Finally, for each die in the network, we consider the average clustering coefficient and it gives rise to four more features. Table I lists all these features.

**Regression Learner:** The quality of our optimization methodology depends on the accuracy of the evaluation function $E$. We can employ any regression learning algorithm e.g. $k$ nearest-neighbor ($k$-*NN*), linear regression (*LR*), support vector regression (*SVR*), and regression tree (*RT*). However, a regression learner that is non-linear, fast in terms of training time and prediction time will improve the effectiveness of the STAGE algorithm. Therefore, analytically, the regression tree learner suits our needs the best.

Our training data consists of a set of input-output pairs $\{(x_i, y_i)\}_{i=1}^{n}$, where each $x_i \in R^m$ is a feature vector and $y_i \in R$ is the corresponding output. The regression tree learning algorithm tries to learn a function $E$ in the form of tree (a set of if-then rules) to minimize the deviation of the predicted output $E(x_i)$ from the correct output $y_i$. The key idea in regression tree learning is to recursively-partition the input space (as in hierarchical clustering) until we find regions that have very similar output values. The recursive partitioning is represented as a tree, where leaves correspond to the cells of the partition. Each leaf is assigned the sample mean of all the output variables in that cell as its prediction. During testing, we find the cell of the partition that input $x$ belongs to through a series of comparison questions on the features, and return the prediction associated with that cell. Regression trees also allow us to identify the features that are important in making predictions.

We employed the WEKA machine-learning toolkit [35] to train regression trees over training set $Z$, and tune the hyper-parameters using validation data.

## V. SPARE-VERTICAL LINK (sVL) ALLOCATION TO IMPROVE THE RELIABILITY OF 3D NoC

The anticipated performance gain of 3D NoC-enabled many-core chips can be compromised due to potential failures of the through-silicon-vias (TSVs) that are mainly used as vertical interconnects in a 3D IC. Workload induced stress is one of the main reasons for the failure of TSV-based VLs in 3D IC. Stress increases the resistance of the TSVs, which leads to different mean-time-to-failure (MTTF) for different TSV based VLs [4][5]. As a result, the performance of the 3D NoC degrades over time, leading to eventual failure of the chip. Therefore, we consider the allocation of spare-VLs as a way to improve the reliability of the 3D NoC.

In this section, we first discuss how the workload induced stress affects the lifetime of TSV-enabled VLs, and then we present the spare VL (sVL) allocation problem formulation and subsequent analysis.

### A. Preliminaries on TSV Reliabilites

To model the failures of vertical links (VLs), we considered the workload-induced stress for TSV-enabled VLs and consequently, explored the effects of resistance increase. In the presence of high current-induced stress, the resistance of the TSVs can increase thereby causing significant delay, and it may eventually compromise the timing specification of the design. From the 3D NoC perspective, depending on the overall interconnection architecture and application characteristics, the traffic densities of the VLs vary. As a result, the workload-induced stress for the TSVs exhibit a non-homogeneous distribution. From the physical perspective, higher workload makes the electro-migration effects more pronounced. With the increase in utilization of TSVs, the resistance and delay of the TSVs also increase and at a certain point, the delay increases beyond the acceptable limit. This scenario can be considered as the failure of TSV and the corresponding time is termed as the mean-time-to-failure (MTTF). In general, 10% increase in resistance can be considered as failure of a TSV [5]. The resistance of TSVs is modeled as shown in (1) and (2) below [5]-

$$R(t) - R_0 = A \ln\left(\frac{t}{t_0}\right) \quad (1)$$

where, $A = \frac{\rho_B}{4\pi t_B}$ and $t_0 = \frac{t_{cu}\pi r_{TSV}^2}{\alpha F} \quad (2)$

The parameter $A$ is called the aging coefficient and $t_0$ is the time when a void (created at the TSV-pad junction) becomes greater than the TSV cross section. The parameters $R(t)$ and $R_0$ refer to the resistance values at times $t=t$ and $t=0$, respectively. The other parameters- $\rho_B$, $t_{cu}$, $t_B$, and $r_{TSV}$ denote the TSV barrier resistivity, copper thickness, barrier thickness, and radius of the TSV respectively. The parameter $\alpha F$ denotes the portion of the vacancy flow, which leads to electro-migration and generates the void under the TSV. Note that the parameter $t$ in the above equations is the active TSV utilization time. Hence, to consider work load induced stress, the active TSV utilization factor (VL utilization factor to be more specific) can

---

**Algorithm 2: Greedy Spare-VL Allocation**

1: **Input:** $F$ = set of $m$ functional VLs,
   $n$ = budget for spare-VLs,
2: **Output:** $S$, the best set of $n$ functional VLs that gets spares
3: **Initialization:** initialize solution set $S = \emptyset$
4: **for** each greedy step =1 to $n$
5:   **for** each choice $x \in F$
6:     value $(x)$ = ***simulator_call*** $(S \cup x)$
7:   **end for**
8:   $x^* = \arg\max_{x \in F} value(x)$
9:   $S = S \cup x^*$  // Functional VL $x^*$ gets spare
10:  $F = F \setminus x^*$  // $x^*$ is removed from $F$
11: **end for**
12: **return** $S$

*\*simulator_call* is a procedure that calculates and returns the network performance and lifetime for a given NoC configuration, benchmark suite, and routing algorithm through extensive experiments.

---

be multiplied with the parameter $t$ to determine the application stress for the TSV-enabled VLs.

### B. Spare VL Allocation Problem

Given a set of $m$ functional VLs $F$ and budget size of spare-VLs $n$ ($n > 0$, $n \ll m$), we want to select the subset of $n$ functional VLs out of $m$ those when provided with *one* spare-VL each will maximize the reliability (lifetime) of the 3D NoC. We can experimentally compute the quality of a given sVL allocation solution by running a simulation. This is an instance of a combinatorial optimization problem with an unknown cost function, where the quality of a given solution can be computed only by making a simulator call. Here, the term 'solution' refers to a particular 3D NoC configuration incorporated with spare-VLs for $n$ functional VLs.

### C. Computational Challenges

The main challenge here is that we have a huge number of possible solutions or NoC configurations ($\binom{m}{n}$) to allocate spare-VLs among the functional links. A naive approach is to enumerate all possible solutions; compute the quality of each solution via simulator call; and pick the best solution. However, the simulator call is expensive in terms of both time and memory requirements. Hence, this exhaustive search approach to quantify the performance and lifetime of each of the candidate NoC configuration is infeasible for practical purposes.

### D. State Space Search Formulation

We solve the sVL allocation problem using a state-space search formulation, where the simulations guide the search process. Each state in our search space is a particular NoC configuration allocated with spare-VLs and consists of a set $S \subseteq F$, where $S$ is a partial or complete solution. Our search space is a 3-tuple $<I, A, T>$, where $I$ is the initial state function that returns the initial search state $S=\emptyset$ meaning solution set is empty; $A$ is a finite set of actions (or search operators) corresponding to growing the partial solution $S$ by one element from $F \setminus S$; and $T$ is the terminal state predicate that maps search nodes to $\{1, 0\}$ indicating whether the node is a terminal or not. Each terminal state in the search space corresponds to a complete solution ($|S| = n$, where $|S|$ denotes the total number of candidates of $S$), while non-terminal states correspond to a partial solution ($|S| < n$). Thus, the decision process for constructing a complete solution corresponds to selecting a sequence of actions leading from the initial state (none of the spare-VLs are allocated) to a terminal state (all the $n$ spare-VLs are allocated). In principle, we can employ any heuristic search procedure (e.g., greedy, beam search) guided by simulations.

*E. Greedy Search for Spare-VL Allocation*

This is the simplest search procedure. We start with an empty solution set $S$. In each greedy step, we add the spare-VL from $F \setminus S$ to the solution set $S$ that when provided with a spare link, it improves the reliability by maximum amount. We repeat this greedy selection step until $S$ is a complete solution ($|S| = n$). The time complexity of greedy search is $O(m*n - n^2)$ simulator calls. Algorithm 2 provides the pseudo-code for greedy spare allocation.

The greedy search is able to produce highly effective sVL allocation that can significantly improve the reliability of the 3D NoC. This effect of greedy sVL allocation was observed through experimental studies as the cost function is unknown and we need to find solutions via simulator calls. The allocation policy to allocate a spare (if spare-VL budget allows) to the first functional VL that fails with a given functional and spare VL-based 3D NoC configuration is highly effective. Intuitively, if we don't allocate spare to the functional VL that is expected to fail first, it will result in a cascade of VL failures reducing the lifetime of the chip drastically.

*F. Domain Knowledge for Sound Pruning*

In 3D NoC enabled many-core chips, some VLs experience heavy traffic and high utilization as the underlying routing algorithm tries to find shortest paths between source and destination cores via these links. As a result, those VLs with high utilization undergo heavier stress, and introduce additional delay in the path and fail more quickly when compared to others. Moreover, this is not an independent phenomenon: one VL failure can decrease the time to failure of a neighboring VL leading to a clustering effect as workload of the neighboring links increase. For example, in Fig 3, we show the traffic densities and the MTTF values of all the VLs for a 64-core and four-layer 3D SWNoC for the CANNEAL benchmark (one of the PARSEC benchmark with highest traffic injection load and skewed traffic). We can see that the traffic densities of VLs 17 to 32 (we call this region as critical region) are significantly higher than that of the others and expectedly, their mean-time-to-failure (MTTF) values are significantly lower.

Our key insight is that for a small budget size $n$ (say less than the number of critical VLs), the spares should be allocated to some of the critical VLs only and there is no benefit for allocating spares to non-critical VLs (chip will fail due the failure of all critical VLs). We can use this domain knowledge to "soundly" prune the search space of possible solutions for the spare-VL allocation problem. Let $H \subseteq F$ correspond to the critical VLs and the total number of critical VLs is $h$ where $h = |H|$. If we consider complete solutions from $H$ only (i.e., subsets of size $n$ from $H$), we can still retain the optimal solution. In other words, we get huge computational savings without losing any accuracy due to sound pruning. For *exhaustive* search, we can consider $\binom{h}{n}$ instead of $\binom{m}{n}$ candidate solutions, where $h <$ $m$. For *greedy* search, we only consider the VLs from $H$ for spare allocation (number of actions $|A|$ will be relatively smaller).

For the rest of the experiments and analysis, we denote the baseline greedy and exhaustive search by **Greedy-Full** and **Exhaustive-Full** respectively. In addition, these techniques enabled with domain knowledge based pruning are named as **Greedy-Restricted** and **Exhaustive-Restricted**.

## VI. EXPERIMENTAL RESULTS AND ANALYSIS

In this section, we first present the achievable performance and energy consumption profiles of our optimized 3D SW NoC architecture. Then we present a detailed reliability analysis in the presence of spare-VL insertion. For this performance evaluation, we consider three metrics: latency, energy consumption, and energy-delay-product (EDP) per message. The EDP is defined as the product of network latency and energy consumption, and it unifies both of them into a single parameter.

*A. Experimental Setup*

To evaluate the performance of different NoCs, we use a cycle-accurate NoC simulator that can simulate any regular or irregular 3D architecture [36]. We consider a Chip Multiprocessor (CMP) consisting of 64 cores and 64 network routers equally partitioned in four layers. In each die, 16 cores are placed in regular interval in a grid pattern. The length of each packet is 64 flits and each flit consists of 32 bits. The routers are synthesized from an RTL level design using TSMC 65-nm CMOS process in Synopsys™ Design Vision. All router ports have a buffer depth of two flits and each router port has four virtual channels in case of irregular NoC. The NoC simulator uses wormhole routing, where the data flits follow the header flits once the router establishes a path. For regular 3D mesh-based NoC, XYZ-dimension order based routing is used. For irregular architectures such as the SW network, the topology-agnostic Adaptive Layered Shortest Path Routing (ALASH) algorithm is adopted [37]. The energy consumption of the network routers, inclusive of the routing strategies, was obtained from the synthesized netlist by running Synopsys™ Prime Power, while the energy dissipated by wireline links was obtained through HSPICE simulations, taking into consideration the length of the wireline links. We consider four SPLASH-2 benchmarks, namely, FFT, RADIX, LU, and WATER [38], and five PARSEC benchmarks, namely, DEDUP, VIPS, FLUIDANIMATE (FLUID), CANNEAL, and BODYTRACK (BT) [39] in this performance evaluation. These benchmarks vary in characteristics from computation intensive to communication intensive in nature and thus are of particular interest in this work.

*B. Performance of the Optimization Algorithm*

In Section IV, we described the details of the STAGE optimization algorithm for designing the 3D SWNoC architecture. Here we first characterize the performance of the optimization algorithm by quantifying various performance metrics of the optimized 3D SWNoC. To evaluate the performance of STAGE algorithm, we compare it with the well-known combinatorial optimization algorithm, viz., simulated annealing (SA) [40] and genetic algorithm (GA) [41] based optimization. We evaluate the performance in terms of both the quality of solution and their individual (for both SA and GA) convergence time.

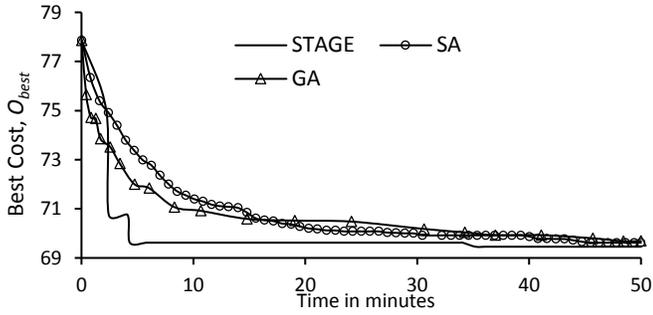

Fig. 4: Performance comparison among the machine-learning-based optimization algorithm (STAGE), the Simulated Annealing (SA) and the Genetic algorithm (GA)

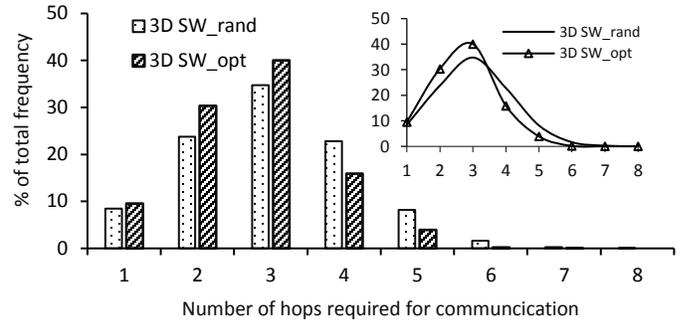

Fig. 5: Effect of optimization algorithm on weighted communication features.

*1) STAGE vs. SA and GA*

We create the initial network following the power law distribution shown in Section IV, where long-range links are placed randomly. Our goal is to find an optimized network starting from this random SW network. We call this initial NoC architecture as **3D SW_rand**. Fig. 4 shows the communication cost of the optimized network from the STAGE and SA algorithm as a function of time.

To compare the performance of these two optimization algorithms, we consider two parameters, viz., the quality of the solution and the convergence time. To make the comparison fair, we consider the same NoC configuration and apply both STAGE and SA algorithm to optimize it. We used a machine configured with Intel Core i7-4700MQ processor and 8 GB RAM running at a clock frequency of 2.4 GHz.

Fig. 4 shows the cost of the best solution obtained at any particular time for SA, GA and STAGE. We consider the best explored cost, $O_{best}$, as the quality of the optimization algorithm. It is evident that STAGE reaches $O_{best}$ very fast (within 5 minutes). During the optimization process, the learned function $E$ predicts an initial network configuration to start the local search procedure that can lead to lower communication cost ($O$). During the initial exploration phase, the error-rate is non-monotonic and high. After a few iterations the prediction error reduces to less than 1%, and after 20 iterations, the error is almost zero (0.05%). The prediction error remained more or less the same for all the subsequent iterations. These results indicate the effectiveness of our network features $\phi$ and the regression-learning algorithm. Note that the best $O$-value decreases monotonically as the set of explored designs increases over the iterations. We also ran the same experiment with the *γ-Greedy* starting state distribution as mentioned above. However, the communication cost $O$ and the prediction error have similar characteristics as the random distribution for the benchmarks and the system size considered in this work. Therefore, we present and discuss our results with a random starting-state distribution.

It is also seen that, both the SA and GA show similar cost function exploration characteristics. Both of them reach $O_{best}$ more gradually compared to STAGE, and even after 50 minutes their respective $O_{best}$ does not reach the same solution as STAGE. It should be noted that we have to optimize the link locations for various applications. Hence, this additional time needed by SA will be a significant overhead when we have to optimize and reconfigure the 3D SWNoC in the field. It should be noted that the final link distribution of the optimized 3D SWNoC is the same for SA, GA and STAGE. However, as shown in Fig. 4 the benefit of STAGE over SA and GA mainly comes from the much faster convergence time. Hence, we can conclude that STAGE algorithm is more efficient in designing an optimized 3D SWNoC with better performance. We denote the final optimized NoC as **3D SW_opt.**

*2) Characteristics of the Design: Random vs. Optimized*

Now we investigate why the STAGE based optimization algorithm is suitable for developing energy-efficient NoC architectures. In Section IV, we described the details of the feature definition ($\phi$), to represent each network. So, we will explore how the design features change before and after the optimization process. Here we specifically consider the role of the weighted communication feature mentioned in Section IV. Fig. 5 shows the weighted communication feature, which reveals the percentage of total communication that is constrained between two nodes separated by $k$ hops ($k \geq 1$). Careful observation of Fig. 5 shows that for 3D SW_opt, the traffic constrained within one, two, and three hop increases compared to 3D SW_rand. Moreover, the amount of traffic that has to traverse beyond three hops decreases.

Hence, the inter-node communication that takes place in less than three hops becomes more frequent. Since the average hop count of the optimized network is calculated to be 2.94, any communication below this average hop count can be considered to be efficient. Essentially, the optimized network becomes more efficient for the same objective function.

The inset in Fig. 5 shows the percentage of communication versus the number of hops, where the area under the curve denotes the weighted communication feature mentioned in Section IV. We can see that the 3D SW_opt curve shifts towards the left, which means that on an average, any message in the optimized network traverses less hops compared to the initial network. Hence, it spends less time inside the network and occupies less network resources. Therefore, the STAGE-based optimization algorithm guides the search to converge to an efficient architecture.

*C. Determination of Connectivity Parameter, α*

As discussed in section III, the connectivity parameter of the small-world network affects the overall performance. Hence, in this section, we experimentally determine the connectivity parameter. For this purpose, we undertook network optimization by varying the connectivity parameter, $α$ for 9 benchmarks considered in this work. To determine $α$, we considered a 64-core system in 4 dies with equal number of cores in each layer. To show how the performance of SWNoC varies with connectivity parameter α, Figs. 6(a)-(b) show the EDP profile for individual benchmarks and their average respectively. From these figure, it is seen that the lowest value for EDP is achieved when the connectivity parameter equals to 2.4. Hence, we choose the value of connectivity parameter to be 2.4 while designing the 3D SWNoC. Hence, for the subsequent experiment and performance comparison, we use this value for the connectivity parameter, α, to develop 3D SWNoC.

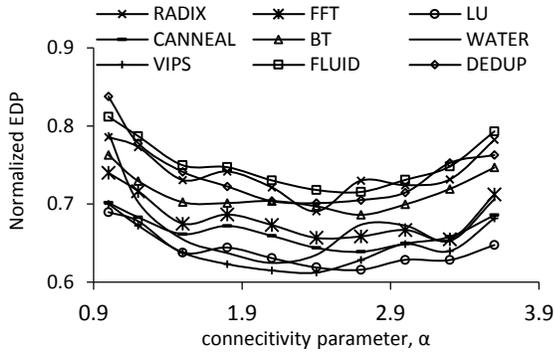

Fig. 6(a): EDP profile with connectivity parameter variation

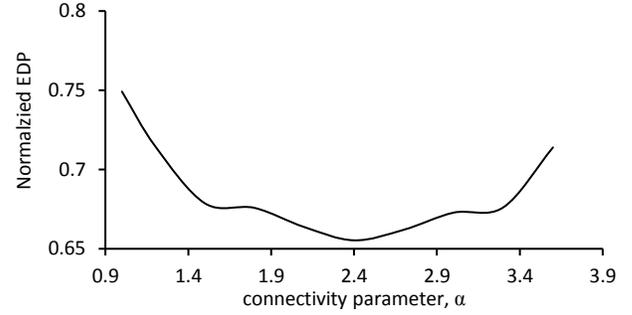

Fig. 6(b): Average EDP profile for 9 benchmarks with connectivity parameter variation

### D. Performance of 3D SWNoC Compared to Other 3D NoCs

In this section, we compare the performance of 3D SWNoC with several existing 3D NoC architectures. For the comparative performance evaluation, we consider 3D MESH and two recently proposed irregular 3D NoCs, namely, *mrrm* and *rrrr* [42]. Both the *mrrm* and *rrrr* NoCs have point-to-point vertical connections as in 3D MESH and 3D SW. However, their die-level planar connection pattern varies. For *rrrr*, all the four dies have randomly connected interconnection patterns. On the other hand, *mrrm* has random connection patterns in the middle two dies whereas the first and the fourth dies follow mesh-based regular connectivity. To build *mrrm* and *rrrr*, we follow the method suggested in [42] and keep the number of links equal to that of 3D MESH and 3D SW. All the performance metric values are normalized with respect to the 3D MESH.

In addition, to show the effect of the optimization algorithm, we evaluate and compare the performances of the optimized NoC architecture with un-optimized counterpart marked as 3D SW_opt and 3D SW_rand respectively.

*1) Network Latency*

Fig. 7(a) demonstrates the normalized network latency of both the 3D SW_rand and 3D SW_opt NoC compared with other existing 3D NoCs. The optimization improves the network latency on an average of 3% over the un-optimized version, and 5.5% over the conventional 3D MESH. The optimization process redistributes the links among the cores such that cores that have to frequently communicate with each other are either directly connected or need to traverse a small number of hops. This results in reduced average hop count and weighted communication for 3D SW_opt NoC.

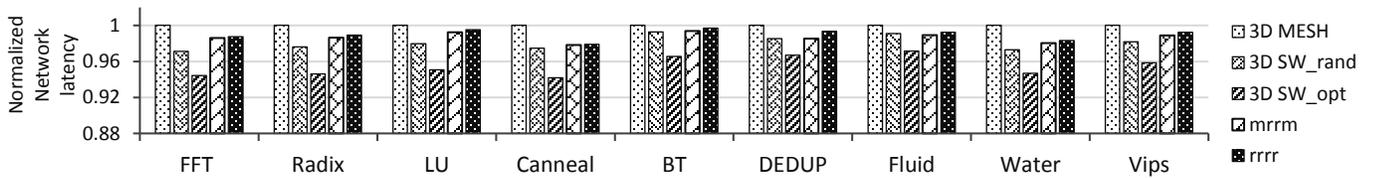

Fig. 7(a): Normalized network latency of 3D SWNoC compared with other 3D NoCs

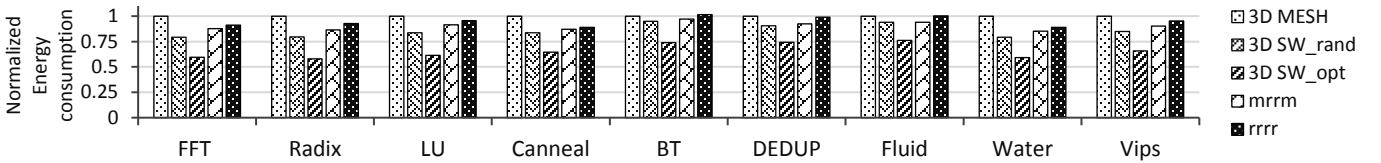

Fig. 7(b): Normalized energy consumption per message of 3D SWNoC compared with other 3D NoCs

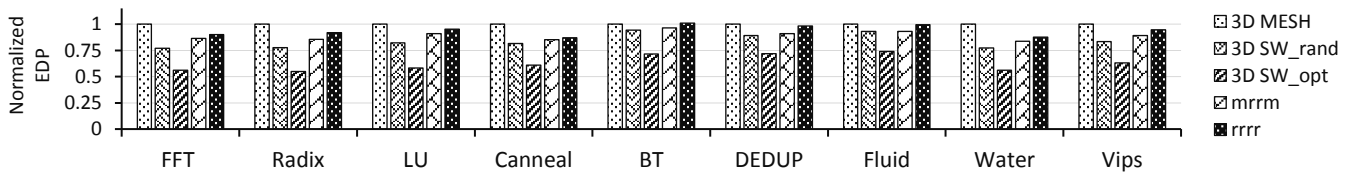

Fig. 7(c): Normalized energy-delay-product (EDP) of 3D SWNoC compared with other 3D NoCs

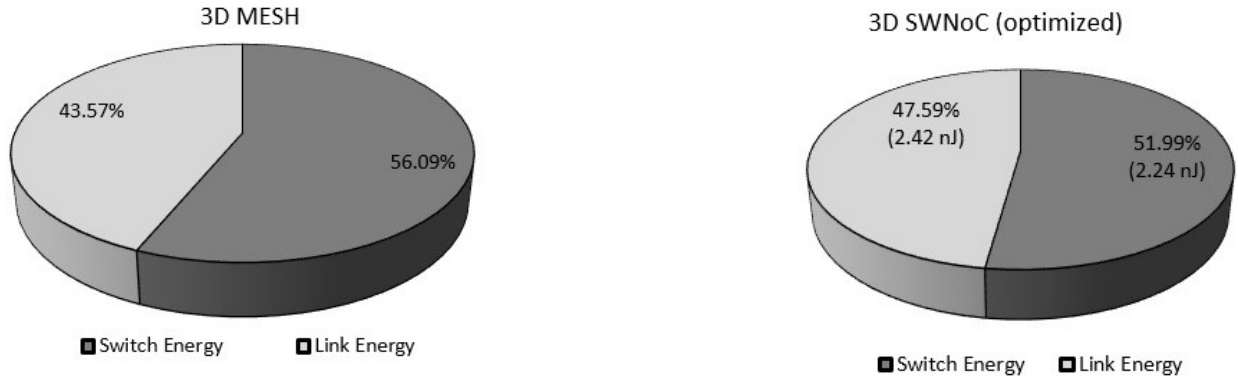

Fig. 8(a): Breakdown of normalized energy consumption profile (energy per message) for 3D MESH and 3D SWNoC (optimized). For 3D SWNoC, on average we achieve 35% energy saving compared to 3D MESH. Large reduction of switch energy (router) is achieved due to reduction in hop count and task remapping to enable small-world network concept. In building 3D SWNoC, highly communicating cores are placed on top of each other or adjacent to each other.

In addition, it is seen that among all the NoCs, 3D MESH and 3D SW_opt exhibit the highest and the lowest latency respectively. The two other architectures namely *mrrm* and *rrrr* perform somewhere in the middle. As in the case of 3D SW NoC, both *mrrm* and *rrrr* have irregularities in the horizontal planes. However, the number and the length of the links are not optimized for these architectures. For *rrrr*, the link distribution has large number of long-range links that help communication among long-distant cores at the expense of near-by communication. In the case of 3D SW_opt NoC, the link distribution follows the power law and the connection pattern is optimized to facilitate both the nearby and long-range communications.

The *mrrm* architecture maintains the link distribution in between *rrrr* and 3D SW_opt NoC. Hence, its network latency lies in between *rrrr* and 3D SW_opt. Finally, 3D MESH NoC suffers from higher average hop count compared to other 3D architectures due to multi-hop communication pattern; hence, it suffers from the highest network latency. Table II lists the communication costs and average hop counts for all these NoCs. As expected, 3D SW_opt and 3D MESH exhibit the lowest and highest communication cost and hop count respectively, whereas *mrrm* and *rrrr* reside in between these two. The effect of these communication costs is eventually reflected in the latency characteristics.

TABLE II.   COMPARISON OF AVERAGE HOP COUNT AND COMMUNICATION COST OF 3D NoC ARCHITECTURES

| NoC architecture | Avg. hop count | Communication cost, $O$ |
|---|---|---|
| 3D SW | 2.94 | 69.45 |
| *mrrm* | 3.07 | 76.15 |
| *rrrr* | 3.03 | 76.34 |
| 3D MESH | 3.81 | 83.47 |

*2) Energy Consumption*

Energy consumption per message depends on the amount of energy consumed by the router as well as the planar links and VLs. The STAGE-based optimization algorithm reduces the average hop count and communication cost by optimizing the objective function $O$ specified in Equation 1. As a result, both the router and link energy consumption are minimized. Fig. 7(b) plots the energy consumption profile of the 3D SWNoC before and after optimization along with the profile for other 3D NoCs. All the energy values are normalized with respect to the corresponding values for the 3D MESH. On an average, the 3D SW_opt NoC shows 33% and 17% energy consumption improvement over the 3D MESH and 3D SW_rand respectively. Fig. 5 helps us in understanding the reasons behind the improvement in energy consumption. The area under the 3D SW_opt curve is less than that of the un-optimized counterpart. Hence, 3D SW_opt reduces the utilization of network resources for any message. As a result, both the router and link energy decrease and the overall energy consumption profile improves.

Fig. 7(b) also demonstrates that among all other NoCs, 3D MESH has the highest energy consumption followed by *mrrm, rrrr,* and 3D SW_opt NoC. Higher network latency of any NoC increases the utilization of network resources and hence, higher energy consumption per message. For 3D MESH, the router energy consumption is significantly higher due to multi hop communication, so it performs the worst among all of them. The *mrrm* and *rrrr* NoCs are capable of reducing the router energy consumption compared to mesh and performs better than 3D MESH. However, due to their random link distribution, they suffer from higher communication cost and average hop count compared to the optimized SW NoC. Hence, they consume more link energy and router energy. With the least communication cost, 3D SW_opt consumes the lowest energy possible among all these architectures.

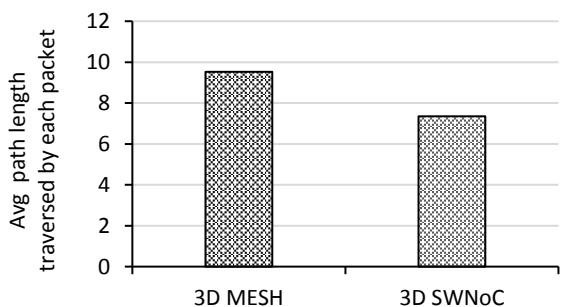

Fig. 8(b): Average path length for every message for 3D MESH and SWNoC

To explain the energy consumption profile of 3D SWNoC, we also breakdown total consumption into two major parts, viz. switch and link energy. Fig 8(a) shows the switch and link energy for 3D MESH and 3D SWNoC. From these experimental results, we can see that the switch energy dominates over the link energy for overall energy consumption (on average ~56% router energy and ~34% link energy). We also show the switch and link energy distribution both for 3D MESH and SWNoC. It is clear that 3D SWNoC saves both switch and link powers. The switch power reduction comes from the lower hop count of small-world networks compared to a MESH. On the average each message traverses lesser number of switches and links. Also, Fig. 8(b) shows the average link length traversed by each message for the 3D MESH and the 3D SWNoC. As the average path length is reduced significantly for 3D SWNoC compared to 3D MESH, the overall energy consumption profile also improves for SWNoC architecture.

*3) Energy-delay-Product (EDP)*

The energy-delay-product is directly affected by the network latency and energy consumption. The architecture that performs best in terms of latency and energy consumption is expected to have lower EDP compared to the EDP of other 3D NoCs. Fig. 7(c) presents the EDP profile of both the un-optimized and optimized 3D SWNoCs along with other 3D NoCs. From the EDP profile, we observe that the average EDP of 3D SW_opt NoC is reduced by approximately 35% and 19% compared to 3D MESH and 3D SW_rand respectively. In addition, among all the other 3D NoCs, 3D SW_opt NoC has expectedly the lowest EDP profile followed by *mrrm*, *rrrr*, and 3D MESH. For all the benchmarks, 3D SW_opt shows the best EDP improvement of 43% for RADIX. The improvement in the network latency and energy consumption of the 3D SW_opt has resulted in similar performance improvement for EDP profile over the other 3D NoCs.

From the above results and analysis, we can conclude that 3D SW_opt NoC performs better than all other considered NoCs in terms of network latency, energy consumption, and EDP. Hence, for the rest of the experiments, we consider this optimized 3D SW architecture and denote it by 3D SWNoC for simplicity.

*E. Performance of 3D NoCs in Presence of Link Failures*

In this section, we analyze the robustness of the 3D SWNoC architecture under VL failure. The reason behind studying the scenario of VL failures is that despite the recent advancements in TSV technology, TSVs are still subject to failure due to voids, cracks, and misalignment [4][43]. In addition, TSVs also face the wear-out problem due to stress arising from potentially high workload. The imbalance in workload among different TSV-based VLs in the NoC also creates wide variation in TSV mean-time-to-failure (MTTF), where some VLs fail early compared to others. Due to all these reasons, if the TSV-based VLs fail, then the EDP and network latency of 3D NoC increases and in the worst case, the corresponding NoC may contain disjoint source-destination pairs in the network. As a result, the performance of the 3D NoC degrades over time.

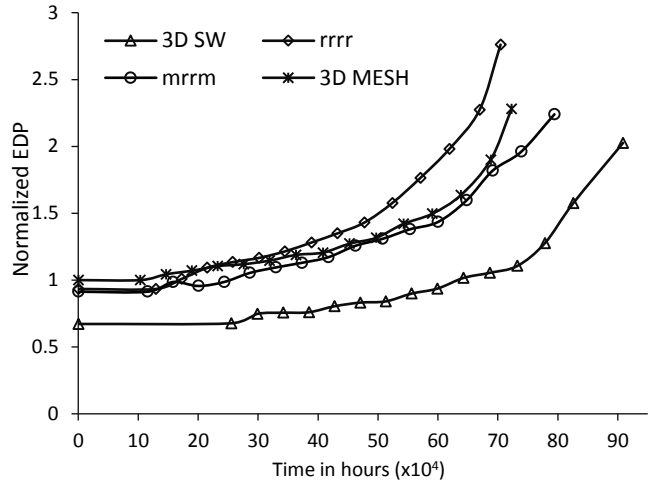

Fig. 9: Normalized EDP profile of 3D NoCs with workload induced vertical link failure scenario for CANNEAL benchmark. The EDP is normalized with respect to fault free 3D MESH at t=0.

Fig. 9 demonstrates the EDP profile of 3D MESH, *mrrm*, *rrrr*, and 3D SWNoC with workload induced VL failure scenario with time for the CANNEAL benchmark (for up to 15 faults). Just like the previous plots, all the EDP values are normalized with respect to fault free 3D MESH at t=0. From the figure, we can see that the EDP values of all the 3D NoCs increase with time as the VLs fail progressively. Among all the NoCs, 3D SWNoC shows the lowest EDP value for any particular period of time and the rate of increase in EDP is also lower than the other NoCs. As a result, 3D SWNoC is expected to have longer lifetimes relative to other NoCs. Note that 3D SWNoC is inherently robust against link failure and its average hop count increases only marginally in the presence of link failures due to the small-world nature of the overall connectivity [28][29]. As a result, it shows better robustness and EDP profile in comparison to other NoCs.

To address this time-dependent failure of VLs and the EDP performance degradation, we propose to incorporate spare-VL allocation to improve the lifetime of the 3D NoC. As 3D SWNoC is inherently more robust than other NoCs, we focus on allocating spare-VLs to 3D SWNoC and analyze its performance with such an allocation.

VII. PERFORMANCE OF 3D NOCS WITH SVL ALLOCATION

In this section, we analyze the sVL allocation methodology and its effects on lifetime and overall reliability of the 3D NoC. First, we characterize the performance of Greedy and Exhaustive search algorithms for sVL allocation. Next, we analyze how the sVL allocation improves the EDP profile and hence, the reliability and lifetime of 3D NoCs.

To quantify the effects of sVL allocation, we first define the lifetime of the 3D SWNoC. Whenever any functional VL fails in a 3D SWNoC, the average hop count increases and hence, the network latency and EDP increase as well. Eventually, the

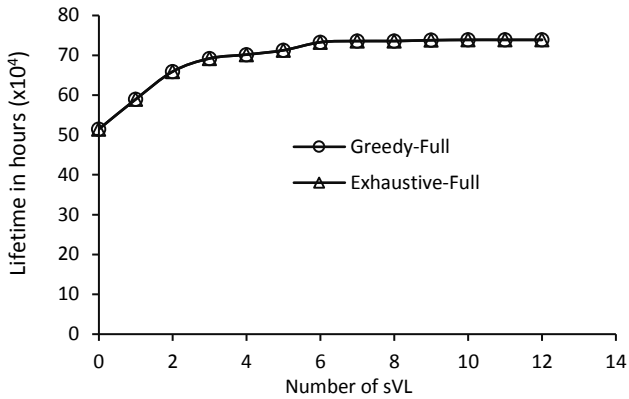

Fig. 10(a): Lifetime of the 3D SWNoC as a function of the number of sVL for the CANNEAL benchmark: Greedy-Full vs. Exhaustive-Full

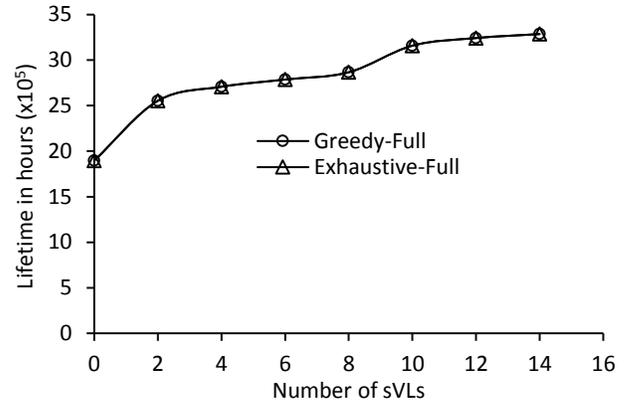

Fig. 10 (b): Lifetime of the 3D SWNoC as a function of the number of sVL for DEDUP benchmark: Greedy-Full vs. Exhaustive-Full

EDP of the 3D SWNoC may be higher than fault free 3D MESH, where it can no longer be considered as an efficient architecture. At this point, the 3D SWNoC loses its architectural advantages over a 3D MESH. We consider the time required to reach this configuration as the *lifetime* of the 3D SWNoC.

*A. Greedy vs. Exhaustive Search for sVL Allocation*

For the sVL allocation, we explored two different allocation algorithms as explained in section IV: Greedy search and Exhaustive search. The allocation algorithms are named as Greedy-Full and Exhaustive-Full (The suffix 'Full' is added to differentiate the algorithms with domain knowledge based pruning, which we will introduce later). For brevity, we show results for three representative benchmarks with varying traffic patterns, viz., CANNEAL, DEDUP, and VIPS. These benchmarks are chosen because they have a wide variation in message injection rates, e.g., high (CANNEAL), medium (DEDUP), and low (VIPS). Fig. 10 (a)-(c) plot the lifetime of the 3D SWNoC with sVL allocation using Greedy-Full and Exhaustive-Full algorithms for different number of spare-VLs for the CANNEAL, DEDUP and VIPS benchmarks respectively. From these figures, we can see that both Greedy-Full and Exhaustive-Full sVL allocation algorithms achieve the same lifetime for the 3D SWNoC. Note that, greedy search (as expected) takes significantly less computation time to produce the solution when compared to exhaustive search. To understand the reason for this behavior, we first need to understand the details of the sVL allocation procedure. To be more specific, the VL failure sequence and its effects on NoC performance need to be explored.

If any functional VL fails, then the workload of this particular VL negatively affects the other neighboring VLs and as a result, the EDP increases rapidly. Consequently, allocation of sVLs to the functional VL, which fails first, is expected to minimize the NoC performance penalty. If sVL is allocated without following the VL failure sequence, then the allocation effect may not be visible on both the EDP profile and lifetime at all.

To explain this behavior in more detail, we consider the case of the CANNEAL benchmark for the 3D SWNoC and number the 48 VLs serially starting from 1 to 48 for a 64 core system (as shown in Fig. 3). For 8 spare-VLs, the sVL allocation solution from exhaustive search corresponds to assigning spares to functional VLs numbered 26, 22, 27, 10, 42, 43, 7, and

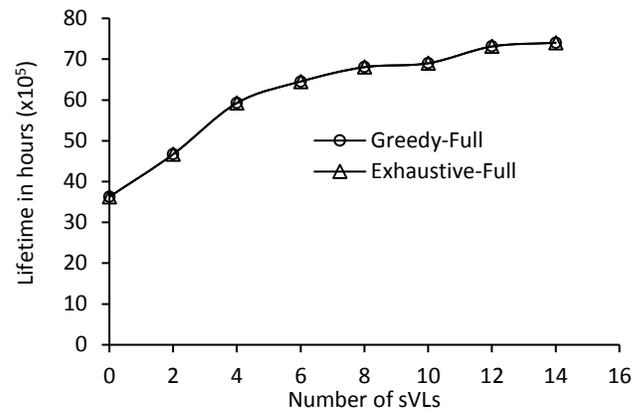

Fig. 10 (c): Lifetime of the 3D SWNoC as a function of the number of sVL for the VIPS benchmark: Greedy-Full vs. Exhaustive-Full

6. Somewhat surprisingly, greedy search also produced the same sVL allocation solution. Our experimental analysis showed that the greedy search produces sVL allocation solutions that can significantly improve the reliability of the 3D NoC. The allocation policy to allocate a spare (if the sVL budget allows) to the first functional VL that fails with a given functional and spare VL-based 3D NoC configuration is highly effective. Intuitively, if we do not allocate spare to the functional VL that is expected to fail first, we will be faced with a cascade of VL failures, which will reduce the lifetime of the chip drastically. For example, the VL failure sequence without any spares allocated is 26, 22, 27, 10, 32, 30, 25, 18 and so on. Greedy search allocates the first spare to functional VL 26. The VL failure sequence after assigning spare to VL 26 is 22, 26, 27, 23, 32, 30, 31, 25, 18 and so on. Greedy search allocates the second spare to functional VL 22. Continuing this policy, greedy search assigns spares to the same set of functional VLs as done by the exhaustive search. We found this behavior to be consistent across all the benchmarks.

*B. Domain Knowledge for Pruning the Search Space*

The time to compute the sVL allocation solution grows as the number of functional VLs ($m$) and the number of sVLs ($n$) increases for both exhaustive search and greedy search. The time complexities of exhaustive search and greedy search (in terms of the number of simulator calls) are $O(\binom{m}{n})$ and $O(mn - n^2)$, respectively. For example, for a 64 core 3D NoC with $m=48$ and $n=8$, the total solution exploration times for exhaustive and greedy search are $377,348,994q$ and $356q$ respectively (here $q$ corresponds to the computation time of a single simulator call which is currently ~7 min in the current experimental setup using a machine configured with Intel Core i7-4700MQ processor and 8 GB RAM running at a clock frequency of 2.4 GHz). Therefore, our sVL allocation algorithms may not scale for large-scale 3D NoC. We consider using domain knowledge of the workload of different functional VLs to prune the solution space as described in Section V. In 3D NoC, the workload of some fVLs (say critical VLs) is much higher than the others and hence, their failure probabilities are higher too. Intuitively, when the sVLs budget ($n$) is small, it is beneficial to allocate spares to some of the critical VLs only because the chip will fail due to a cascade of critical VL failures.

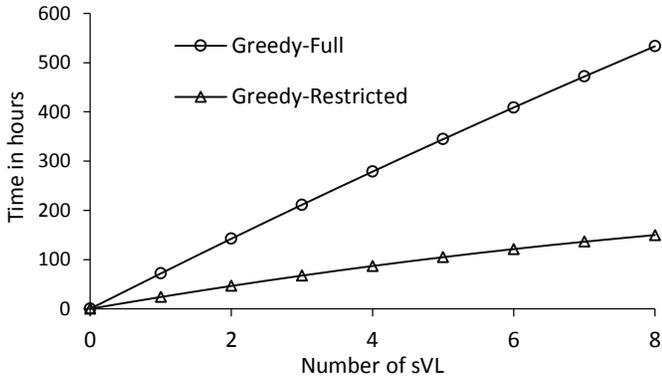

Fig. 11(a): Estimated runtime for different number of sVL: Greedy-Full vs. Greedy-Restricted

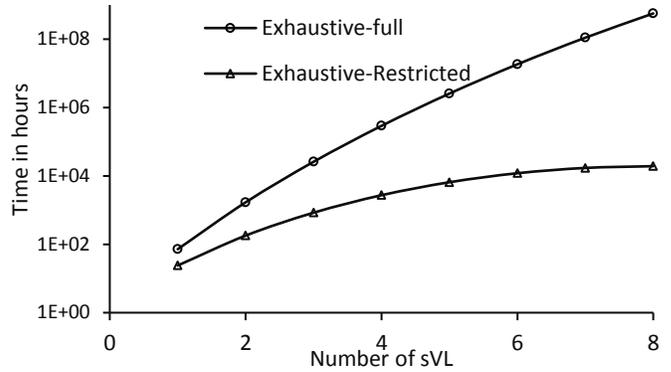

Fig. 11(b): Estimated runtime for different number of sVL: Exhaustive-Full vs. Exhaustive-Restricted

We select a subset of critical VLs (say $H$) out of the $m$ functional VLs that we will consider for allocating spares and prune the remaining ones. Pruning can improve the computational efficiency of solving the sVL allocation problem, but may potentially compromise the accuracy of solutions depending on the amount of pruning. We can consider varying amounts of pruning from $|H|=n$ (only one candidate solution) to $|H|=m$ (no pruning) to trade-off speed and accuracy of producing sVL allocation solutions. A simple pruning strategy to achieve this goal is as follows: rank all the functional VLs according to their workload; select the top-$|H|$ VLs to be considered for spare allocation; prune the remaining $m-|H|$ VLs. We can use both exhaustive search and greedy search to find the solution from this restricted set of candidate solutions. We refer to the exhaustive and Greedy sVL allocation algorithms as Exhaustive-Restricted and Greedy-Restricted, respectively.

Fig 3 shows the traffic densities of all the VLs for a 64-core 3D SWNoC consisting of four planar layers for the CANNEAL benchmark. It can be noted that the traffic densities of some VLs (critical VLs) are significantly higher than that of the others. To identify the critical VLs, we rank VLs according to workload and sort the highest workload ones. In this particular work, we consider 16 critical VLs. This number is chosen considering the worst-case VL failure scenario where all 16 critical VLs are placed in between two adjacent planar dies and if all of them fail together, then the NoC becomes completely unrouteable. Therefore, we prune all the non-critical VLs, a total of 32 out of 48 (other than 16 critical VLs). In other words, $|H|=16$ corresponding to 16 high workload carrying VLs, which is significantly smaller compared to $m=48$ (total number of VLs). We found that with this setting, the search algorithms with pruning (Exhaustive-Restricted and Greedy-Restricted) produce the same sVL allocation solutions as their counterparts without any pruning (Exhaustive-Full and Greedy-Full) for different number of spare-VLs ($n=1$ to any number of upper limit). In other words, we do not lose accuracy due to pruning. We do not show these results for the sake of brevity. The main benefit of pruning is that it will improve the computational efficiency of producing sVL allocation solutions using Exhaustive and Greedy search. As an example, Fig. 11(a) shows the estimated runtime comparison of Greedy-Full and Greedy-Restricted for the different number of sVLs. Similarly, Fig. 11(b) shows the estimated runtime comparison of Exhaustive-Full and Exhaustive-Restricted. We can see that the computational gains are significant due to pruning, but without losing any accuracy.

It is to be mentioned that for comparison purpose of different spare-allocation strategies, the runtimes of the algorithms were converted to equivalent runtime of a single machine (windows based) configured with Intel Core i7-4700MQ processor and 8 GB RAM running at a clock frequency of 2.4 GHz. Hence, the run time of sVL allocation experiments is really large. With a high-speed server cluster and parallel experiments, this runtime reduces significantly. For example, for allocating 16 sVLs among 48 fVLs with Greedy allocation, by running 1000 simulations in parallel in the cluster, it takes less than 1 hour.

For performing exhaustive search experiments for sVL allocation, we employed experimental observations to discard very low-quality solutions. The key observation is that if a functional VL doesn't fail, then it doesn't affect the reliability of the chip. Indeed, less utilized vertical links (tail end of the histogram of utilizations) don't fail before the chip fails (as per the lifetime definition). The 3D SWNoC reaches its lifetime (EDP increases beyond 3D MESH) when 15 to 20 VLs fail. Specifically, we observed that almost 50% of the less utilized VLs never fail. Hence, any sVL solution that includes these less utilized VLs will be of low-quality. Therefore, we can safely discard these low-quality solutions to speed up the experiments without any loss in accuracy. For example, if 50% less utilized VLs are not considered, the worst case runtime for exhaustive search is 315 485 equivalent hours (~7 min per simulation), which was done using the parallel compute cluster in few days (~9 days).

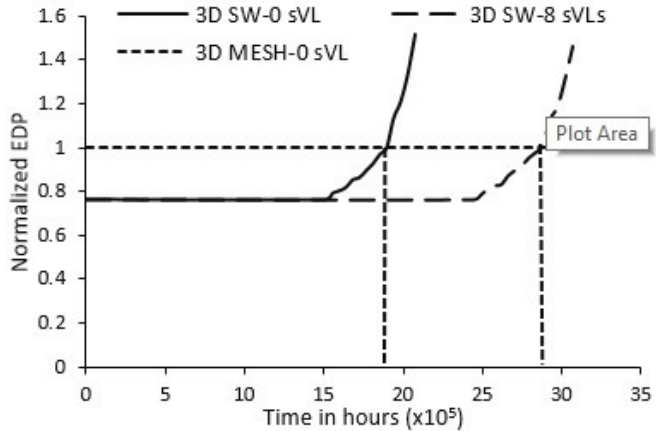

Fig. 12: Lifetime determination algorithm is explained with normalized EDP profile of 3D SWNoC w/ and w/o sVL allocation. As an example, lifetime calculation for 3D SWNoC with DEDUP benchmark has been plotted. The EDP of 3D MESH-0 sVL (dotted line) corresponds to time t=0 and extended only for reference purpose.

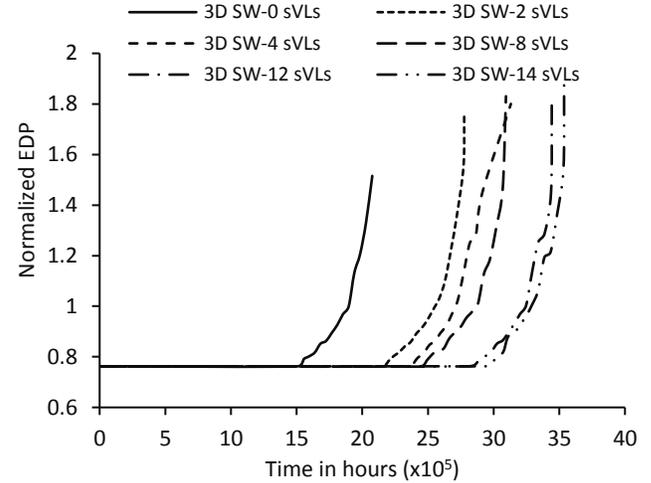

Fig. 13(b): Normalized EDP profile for the 3D SWNoC with different number of spare VLs for the DEDUP benchmark

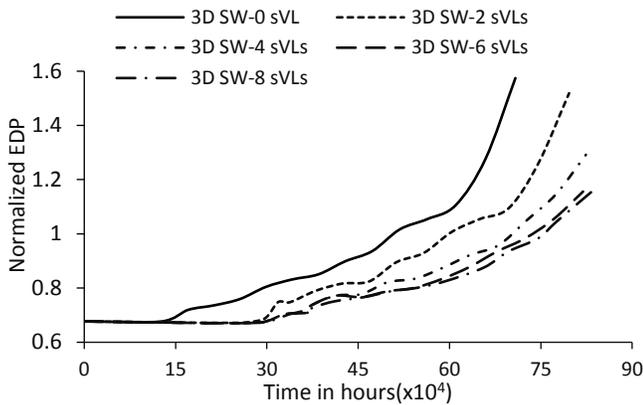

Fig. 13(a): Normalized EDP profile for 3D SWNoC with different number of spare VLs allocation for the CANNEAL benchmark.

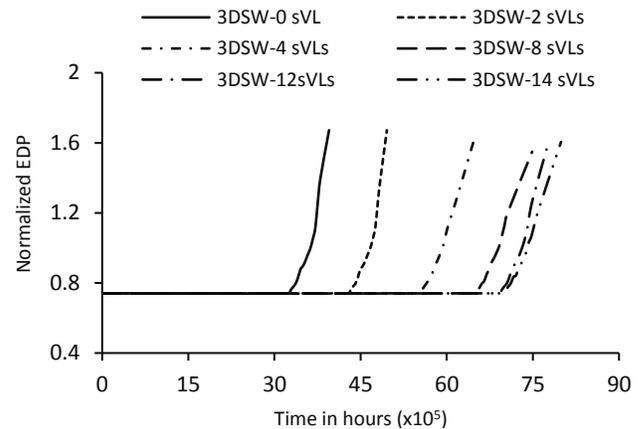

Fig. 13(c): EDP profile for 3D SWNoC with different number of sVL allocation for the VIPS benchmark

## C. Computing the Lifetime of 3D SWNoC with sVL allocation

In this section, we describe the procedure to compute the lifetime of any 3D NoC configuration. For better understanding, we plot the EDP profile of 3D SWNoC with and without spare-VLs incorporated into it, and graphically illustrate how to calculate the lifetime of any 3D NoC.

As defined in the earlier section, the lifetime of any 3D NoC is the time when the EDP value of that particular NoC equals to a certain threshold value. Since the performance requirement for the NoC is application and/or user dependent, the threshold value to compute the lifetime of the 3D NoC will vary.

Fig. 12 illustrates the lifetime computation procedure for a 3D SWNoC incorporated with 8 sVLs for the DEDUP benchmark. This particular configuration is chosen as an example, however, the procedure is applicable for any other 3D NoC and benchmark. For the reference purposes, the EDP profile of the original 3D SWNoC (without any sVL) is also plotted. The EDP of 3D SWNoC is normalized with respect to the EDP value of fault free 3D MESH. To help illustrate the lifetime computation procedure more clearly, a dotted horizontal line is drawn in Fig. 12, which we call the lifetime line. This line corresponds to 100% EDP value for the fault free 3D MESH (at t=0).

We have calculated the lifetimes of 3D SWNoC and 3D SWNoC with 8 sVLs as marked with respective vertical lines. The lifetimes of these NoCs are the projection of the intersection point of the lifetime line and the corresponding EDP profile lines of 3D SWNoC with and without spare-VLs on the time axis. These are marked as $L_{3D\ SW}$ and $L_{3D\ SW\ with\ 8\text{-}sVL}$ in the figure. The lifetime of other 3D NoCs can be calculated in a similar way.

## D. Effects of Spare-VL Allocation on 3D NoC

Whenever a spare-VL is allocated to a functional VL, the sVL carries the traffic when the corresponding functional VL fails. This minimizes the effect of VL failure on 3D NoC performance degradation and essentially helps in maintaining lower EDP value over longer period of time. However, there exists an upper limit for the sVL number, beyond which the advantages of sVL allocation can no longer be pronounced. We call this number as the optimum number of spare-VLs.

Depending on the benchmark and NoC configuration, the optimum number of sVL varies. In this work, we consider 3D SWNoC architecture as the testbed for evaluating the performance of sVL allocation. However, subsequent experiments and analysis are equally applicable for other 3D NoC architectures as well.

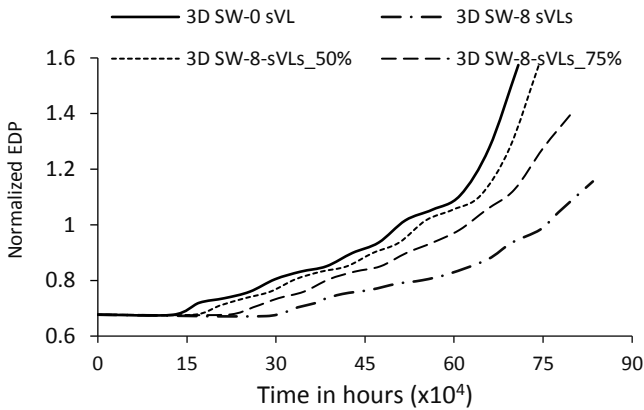

Figure 14(a): Normalized EDP profile of 3D SWNoC for CANNEAL benchmark with 8-sVL allocation. Here, 3D SW-8-sVLs_x% denotes the partial sVL allocation where x denotes the percentage of total TSVs needed to enable one full vertical link

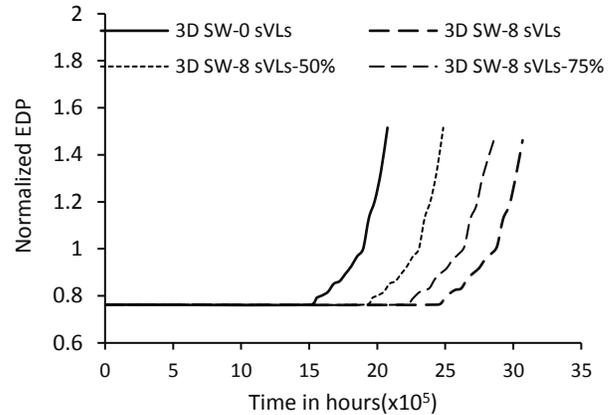

Figure 14(b): Normalized EDP profile of 3D SWNoC for DEDUP benchmark with 8-sVL allocation. Here, 3D SW-8-sVLs_x% denotes the partial sVL allocation where x is the percentage of total TSVs needed to enable one full vertical link

*1) Optimum Number of spare VLs*

In this section, we evaluate the effects of different number of sVLs on the 3D NoC performance. Fig. 13 (a), (b) and (c) demonstrate the normalized EDP of 3D SWNoC with time for CANNEAL, DEDUP, and VIPS benchmarks respectively. Similar to the previous experiments, we have considered these three benchmarks as the representative of high, medium, and low injection benchmarks from the PARSEC and SPLASH-2 suites. All the EDP values are normalized with respect to the EDP of fault free 3D MESH with no sVLs allocated to it at t=0.

From these figures, we can see that the EDP remains unchanged up to a certain point and after that, it increases when the functional VLs start failing. This happens due to the fact that initially no functional VL fails and EDP remains constant up to a certain time. Subsequently, VLs from the critical region (as defined in section V.E) having high traffic density start failing. In such a link failure scenario, the traffic of the failed VLs is carried by the neighboring VLs along with their own traffic. This has two kinds of negative effects. Firstly, the EDP and the network latency of the NoC increases due to a critical link failure. Secondly, the neighboring functional VLs also fail quickly which further degrades the NoC performance. As a result, the EDP increases at a faster rate.

Another interesting result is that as the number of allocated spare-VLs increases, the EDP profile shifts towards the right on the time scale. This implies that the 3D SWNoC with sVL allocation can maintain a particular EDP level for a longer period of time. Expectedly, the lifetime of 3D SWNoC also increases with sVL allocation. In addition, we can see that the difference between the EDP profiles on the time axis decreases gradually as the sVL number increases. For the CANNEAL benchmark, the right-most EDP is found to be for 8 sVLs. We can see that even if we increase the number of sVLs beyond 8 for CANNEAL, the EDP profile doesn't shift to the right anymore. This implies that any further improvement of EDP profile is not possible and we call this scenario as the saturation effect of sVL allocation. Similarly, for 3D SWNoC with

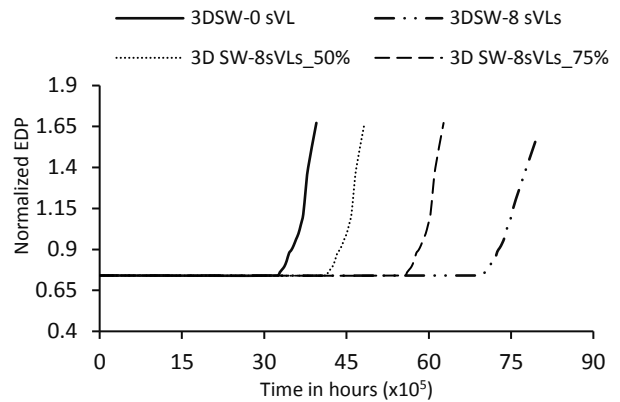

Figure 14(c): Normalized EDP profile of 3D SWNoC for VIPS benchmark with 8-sVL allocation. Here, 3D SW-8-sVLs_x% denotes the partial sVL allocation where x is the percentage of total TSVs needed to enable one full vertical link.

DEDUP and VIPS benchmarks, the EDP profile gets saturated for 14 sVLs.

*2) Performance of 3D SWNoC with Partial sVL-allocation*

In this section, we evaluate the performance of 3D SWNoC with partial sVL allocation. With partial sVL allocation, instead of allocating an sVL (total bundle of TSVs replacing the whole VL), we only allocate some spare TSVs in an fVL and compare its performance with full sVL-allocation explored earlier. For partial sVL allocation, we need to consider the cross-coupling capacitance of the individual TSVs. If we consider a grid-based layout of the TSVs in a bundle, then the centrally located TSVs will have the highest cross coupling. We replace the TSVs that are affected most by the cross coupling in this partial allocation. As a case study, we consider this partial TSV allocation to the critical fVLs only and allocate 50% and 75% of the total TSVs in an fVL. We characterize the performance of this partial TSV allocation with the full sVL allocation.

Figs. 14 (a)-(c) show the EDP profile with time of 3D SWNoC with partial sTSV allocation. In these figures, 3D SW-8sVLs denotes the performance of 3D SWNoC with 8 sVLs allocation (complete bundle allocation) whereas the 3D SW-8-sVLs_x% denotes the performance of 3D SWNoC with individual sTSV allocation within the bundle (VL). For example- *3D SW-8 sVLs_50%* indicate 50% individual TSVs within the bundle (for 8 VLs) have spare TSV allocated to it. From the figures, it is clear that, complete sVL allocation performs better than partial sTSV allocation. As the percentage of spare TSV allocation increases, the EDP profile shifts right on the time scale and lifetime improves consequently. It should be noted that if we allocate 100% sTSVs, then it is equivalent to full-sVL allocation (3D SW-8sVLs in the figure) and achieves the best EDP profile and maximum lifetime for the 3D NoC.

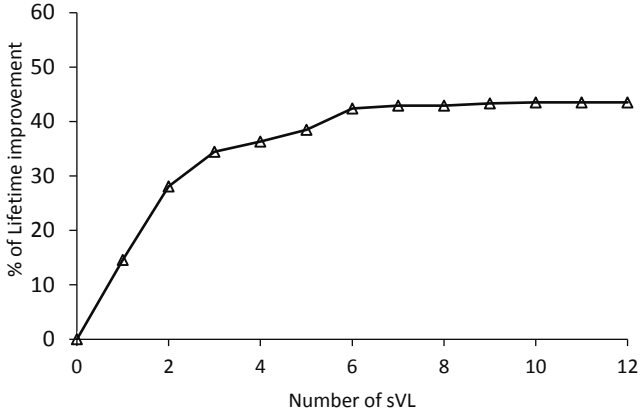

Fig. 15: Effect of sVL allocation on 3D SWNoC for the CANNEAL benchmark. The improvement of lifetime of 3D SWNoC initially increases linearly and saturates beyond 8-sVL allocation. The gain is normalized w.r.t. the initial lifetime of 3D SWNoC at t=0.

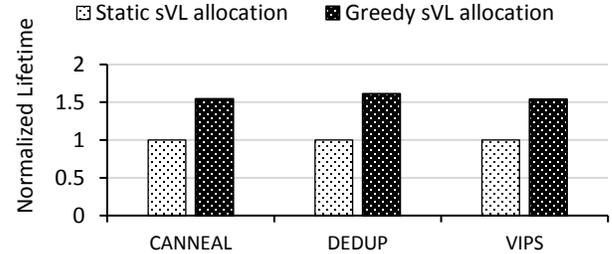

Fig. 16: Lifetime comparison for 3D SWNoC Greedy sVL allocation and static sVL allocation (allocation of sVLs to the most critical VLs)

*3) Saturation of Lifetime Improvement*

In this work, we have considered one-to-one correspondence between sVLs and functional-VLs where any sVL replaces one functional VL regardless of the workload intensity. Allocation of such sVLs increases the traffic carrying capability of the critical VLs and improves the lifetime of the 3D NoC. As an example, Fig. 15 plots the percentage of lifetime improvement of 3D SWNoC for the CANNEAL benchmark with different number of sVLs allocation. Note that similar lifetime improvements are observed for other benchmarks as well.

From the figure, we can see that as the number of allocated sVLs increases, the lifetime of the 3D SWNoC also increases. Initially, the gain of lifetime is almost linear with the number of allocated sVLs and later, the gain increment decreases and improvement saturates after some point. Allocation of sVL increases the combined *lifetime* of the particular VL (consists of sVL and functional VL in this case), which helps to minimize the network latency and EDP degradation due to VL failure.

In general, most critical VLs fail early when compared to the other VLs. If the sVLs are allocated to the critical VLs, then they help in significantly increasing the lifetime of the NoC. However, the lifetime gain saturates as the number of allocated sVL crosses a certain number. This happens due to the fact that the combined lifetime of some critical VLs even with the sVL allocation is shorter than other non-critical VLs. Consequently, even if we allocate sVLs to these non-critical VLs, they do not improve the EDP beyond what is achieved already. We may see benefits if we consider a more general sVL allocation problem, where we consider allocating more than one spare-VL to a functional VL. We defer this for future work. It is important to note that similar effects are also observed for DEDUP and VIPS benchmarks as well (in these cases, the saturation effect was observed for 14 sVLs). However, we have omitted plotting such repetitive results and analysis.

*E. Performance of Static sVL allocation*

In previous sections, we have analyzed the performance of Greedy and Exhaustive sVL allocation mechanisms. In addition, NoC domain knowledge was applied to identify the critical VLs and prune out significant parts of search space for improving the computational efficiency of the algorithms. With this pruning technique, Greedy- and Exhaustive-Restricted approaches explore the fVL failure sequence and allocated the sVLs sequentially following this failure order. However, if we don't explore this fVL failure sequence and naively allocate the sVLs to the most critical fVLs explored at t=0 (only based on the workloads of the fVLs of the 3D SWNoC without any spare allocated to it), then it fails to capture the full advantages offered by the Greedy/Exhaustive sVL allocation algorithms. To explain this, we consider the case, where Greedy algorithm allocates spares to the functional VLs (one or multiple spares to a single fVL) depending on the explored failure sequence at that particular instance. Fig. 16 shows the normalized lifetime of 3D SWNoC with Greedy and static (allocation of sVLs to the most critical ones) sVL allocation for 8 sVLs. In this case also, we have chosen three representative benchmarks with wide variation in characteristics e.g. CANNEAL, DEDUP and VIPS.

From the figure it is seen that, the lifetime of 3D SWNoC with Greedy spare allocation is much higher than static sVL allocation. For CANNEAL, DEDUP and VIPS benchmarks, Greedy spare allocation shows 54%, 61% and 53% higher lifetime respectively compared to the static allocation. This happens due to the fact that in a 3D SWNoC (or any other 3D NoCs), the most critical VL/s may fail several times within the considered time span. Static allocation algorithm only determines the critical fVLs at t=0, and allocates spares without further exploring the sequence at runtime. On the other hand, Greedy method updates the fVL failure sequence at every step and allocates spares accordingly. Hence, Greedy algorithm exploits maximum benefits offered by any possible sequence of sVL allocation and improves the lifetime significantly. Thus Greedy sVL allocation is capable of exploring better solutions compared to static sVL allocation.

*F. Other 3D NoC Reliability Improvement Mechanism*

Careful observation on the above results and analysis reveal that in order to increase the lifetime of the chip, other possible solutions to the TSV aging problem are to redesign "the link timing" and "pipelining at either end of the TSVs". This will enable the TSV-based vertical links to tolerate more delays. However, these types of design techniques are very specific to the allowable timing budget. The circuit parameters need to be modified depending on the varying timing budget. In contrast, we aim to establish a spare link allocation method that does not depend on any specific circuit parameters for improving the reliability of 3D NoCs considering the TSV aging. To this end, the failure criteria for TSVs with 10% delay increase (considered in this work) is merely a reference point that has been adapted from earlier works [4][5]. Any other definition of TSV failure criteria can be adapted for our proposed methodology as well.

In addition, with any other modification in link design, the failure sequence of TSVs explored in this work will remain same for the same set of workloads and NoC architecture. In other words, even with the redesign of TSV-based links or pipelining on either end of TSVs, the qualitative nature of "EDP vs time" profile for 3D SWNoC with spare TSV-allocation remains the same. However, it will shift to the right side in the time axis in Figures 10-12. This will happen due to the fact that the effective failure time of the TSV-based links will increase as they have more time margins. So the qualitative analysis presented in this work with spare TSV allocation will still hold; however, the improvement in chip lifetime may change depending on the link redesign steps or lifetime definition of TSVs. Hence, we conclude that the presented results and discussions on TSV reliabilities and 3D NoC lifetime serve as baseline analyses for future TSV-based design methodologies in 3D NoCs.

## VIII. CONCLUSIONS

We proposed a robust design optimization methodology to improve the energy efficiency of 3D NoC architectures by combining the benefits of SW networks and machine-learning techniques to intelligently explore the design space. We showed that the optimized 3D SWNoC architecture outperforms the existing 3D NoCs. The optimized 3D SW NoC on an average achieves 35% EDP reduction over conventional 3D MESH. We also demonstrated the efficacy and robustness of the 3D SWNoC in presence of non-homogeneous workload induced VL failure. The proposed 3D SWNoC shows better resilience and EDP profile against VL failure at any instant of time compared to state-of-the-art 3D NoCs. We also proposed a spare-VL allocation mechanism to address the performance degradation and lifetime shortening problem due to VL failure. We showed that with a small number of spare-VLs, we could exploit NoC domain knowledge to develop efficient and computationally inexpensive algorithms to explore optimal solution. The proposed spare-VL allocation significantly improve the reliability and lifetime of the 3D NoC.

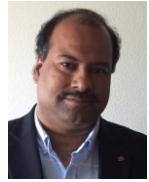
**Partha Pratim Pande** is a Professor and holder of the Boeing Centennial Chair in computer engineering at the school of Electrical Engineering and Computer Science, Washington State University, Pullman, USA.

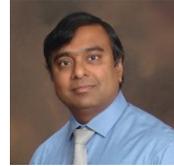
**Krishnendu Chakrabarty** is the William H. Younger Distinguished Professor of Engineering in the Department of Electrical and Computer Engineering at Duke University. He served as the Editor-in-Chief of *IEEE Design & Test of Computers* during 2010-2012 and *ACM Journal on Emerging Technologies in Computing Systems* during 2010-2015. Currently he serves as the Editor-in-Chief of *IEEE Transactions on VLSI Systems*. He is a fellow of both ACM and IEEE.

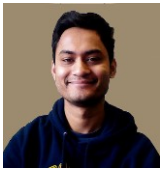
**Sourav Das (**S'14**)** is currently pursuing the Ph.D. degree with the EECS Department, Washington State University, Pullman, WA. His current research interests include low-power network-on-chip design.

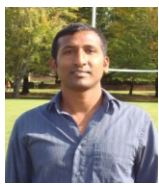
**Janardhan Rao Doppa** received his Ph.D. from Oregon State University. He is an Assistant Professor at Washington State University. His work on structured prediction received an outstanding paper award at the AAAI 2013 conference.